\documentclass[11pt, a4paper]{amsart}
\usepackage[dvips]{epsfig}
\usepackage{amsmath}
\usepackage{amssymb}
\usepackage{amsfonts}
\usepackage{amsthm}
\usepackage{amsbsy}
\usepackage{amsgen}
\usepackage{amscd}
\usepackage{amsopn}
\usepackage{amstext}
\usepackage{amsxtra}
\usepackage{enumerate}
\usepackage{dsfont}
\usepackage{hyperref}
\usepackage{lineno}

\usepackage{times, graphicx}
\usepackage{eso-pic}
\usepackage{lastpage}

\usepackage[paper=a4paper,left=30mm,right=20mm,top=25mm,bottom=30mm]{geometry}
    \newenvironment{dedication}
        {\vspace{6ex}\begin{quotation}\begin{center}\begin{em}}
        {\par\end{em}\end{center}\end{quotation}}

\newtheorem{theorem}{Theorem}[section]

\newtheorem{lemma}[theorem]{Lemma}

\theoremstyle{definition}

\newtheorem{remark}[theorem]{Remark}

\newcommand{\vol}{\operatorname{Vol}}
\newcommand{\prob}{\operatorname{Prob}}

\newcommand {\E} {\mathbb{E}}

\newcommand {\R} {\mathbb{R}}
\newcommand {\N} {\mathbb{N}}
\newcommand {\Z} {\mathbb{Z}}
\newcommand {\C} {\mathbb{C}}

\newcommand {\Cc} {\mathcal{C}}
\newcommand {\Ec} {\mathcal{E}}
\newcommand {\Tc} {\mathcal{T}}

\begin{document}

\title[Topologies of nodal sets]{Topologies of nodal sets of random band limited functions }

\author{Peter Sarnak}
\thanks{Research of P.S. is supported by an NSF grant.}
\address{Department of Mathematics, Princeton University and the Institute for Advanced Study, US}

\author{Igor Wigman}
\thanks{
I.W has received funding from the
European Research Council under the European Union's Seventh
Framework Programme (FP7/2007-2013) / ERC grant agreement
n$^{\text{o}}$ 335141, and the EPSRC grant under the First Grant scheme
(EP/J004529/1)}

\address{Department of Mathematics, King's College London, UK}

\date{\today}

\begin{abstract}
It is shown that the topologies and nestings of the zero and nodal
sets of random (Gaussian) band limited functions have universal
laws of distribution. Qualitative features of the
supports of these distributions are determined. In
particular the results apply to random monochromatic waves and to
random real algebraic hyper-surfaces in projective space.
\end{abstract}

\maketitle

\vspace{-1cm}

\begin{dedication}
{{\bf To Jim Cogdell on his $60$th birthday with admiration}}
\end{dedication}

\vspace{0.5cm}

\section{Introduction}

Nazarov and Sodin (~\cite{NS,So}) have developed some powerful general techniques
to study the zero (``nodal") sets of functions of several variables coming from
Gaussian ensembles. Specifically they show that the number of connected components
of such nodal sets obey an asymptotic law. In ~\cite{Sa} we pointed
out that these may be applied to ovals of a random real plane curve, and in
~\cite{LL} this is extended to real hypersurfaces in $\mathbb{P}^{n}$. In
~\cite{GW} the barrier technique from ~\cite{NS} is used
to show that ``all topologies" occur with positive probability in the
context of real sections of high tensor powers of a holomorphic line
bundle of positive curvature, on a real projective manifold.

In this note we apply these techniques to study the laws of distribution of
the topologies of a random band limited function. Let $(M,g)$
be a compact smooth connected $n$-dimensional Riemannian manifold. Choose
an orthonormal basis $\{ \phi_{j}\}_{j=0}^{\infty}$ of eigenfunctions of its
Laplacian
\begin{equation}
\label{eq:Delta phi+t^2phi = 0}
\Delta \phi_{i}+t_{i}^{2}\phi_{i}=0,
\end{equation}
\begin{equation*}
0=t_{0}<t_{1}\le t_{2} \ldots .
\end{equation*}
Fix $\alpha\in [0,1]$ and denote by $\mathcal{E}_{M,\alpha}(T)$
($T$ a large parameter) the finite dimensional Gaussian ensemble of functions on $M$ given
by
\begin{equation}
\label{eq:f = sum cj phij}
f(x) = \sum\limits_{\alpha T \le t_{j} \le T} c_{j}\phi_{j}(x),
\end{equation}
where $c_{j}$ are independent Gaussian variables of mean $0$ and variance $1$. If $\alpha = 1$,
which is the important case of ``monochromatic"
random functions, we interpret \eqref{eq:f = sum cj phij} as
\begin{equation}
\label{eq:f = sum cj phij mono}
f(x) = \sum\limits_{T-\eta(T) \le t_{j} \le T} c_{j}\phi_{j}(x),
\end{equation}
where $\eta(T) \rightarrow \infty$ with $T$, and $\eta(T)=o(T)$.
The Gaussian ensembles $\mathcal{E}_{M,\alpha}(T)$ are our
$\alpha$-band limited functions, and they do not depend on the choice
of the o.n.b. $\{\phi_{j}\}$. The aim is to study the nodal sets of a typical
$f$ in $\mathcal{E}_{M,\alpha}(T)$ as $T\rightarrow\infty$.

Let $V(f)$ denote the nodal set of $f$, that is
$$V(f) = \{ x:\: f(x)=0\}.$$ For almost all $f$'s in $\mathcal{E}_{M,\alpha}(T)$
with $T$ large, $V(f)$ is a smooth $(n-1)$-dimensional compact manifold. We decompose
$V(f)$ as a disjoint union  $\bigsqcup\limits_{c\in\mathcal{C}(f)}c$ of its connected
components. The set $M\setminus V(f)$ is a disjoint union of connected components
$\bigsqcup\limits_{\omega\in\Omega(f)}\omega$, where each $\omega$ is a smooth
compact $n$-dimensional manifold with smooth boundary. The components $\omega$ in
$\Omega(f)$ are called the nodal domains of $f$. The nesting relations between
the $c$'s and $\omega$'s are captured by the bipartite connected graph
$X(f)$, whose vertices are the points $\omega\in\Omega(f)$ and edges $e$ run
from $\omega$ to $\omega'$ if $\omega$ and $\omega'$ have a (unique!) common
boundary $c\in\mathcal{C}(f)$ (see Figure \ref{fig:nesting tree}). Thus the edges $E(X(f))$ of $X(f)$ correspond to
$\mathcal{C}(f)$.

\begin{figure}[ht]
\centering
\includegraphics[height=50mm]{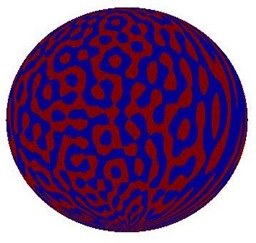}
\caption{A nodal picture of a spherical harmonic. The blue and red are positive
and negative domains respectively, and the nodal set is the interface between these.}
\label{fig:sphere}
\end{figure}

\begin{figure}[ht]
\centering
\includegraphics[height=54mm]{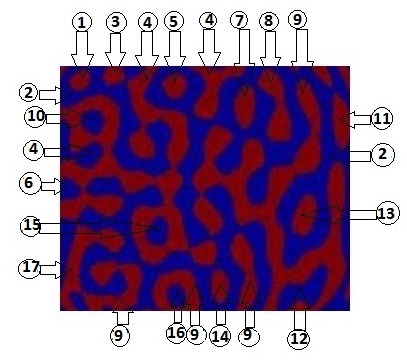}
\includegraphics[height=54mm]{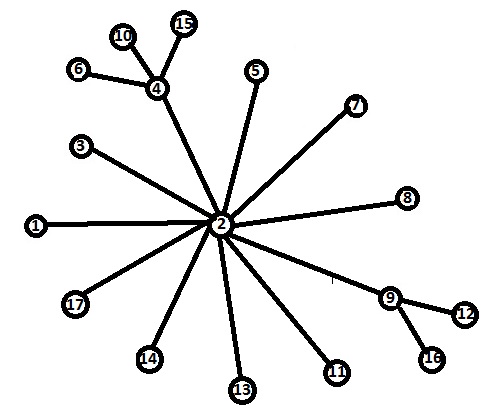}
\caption{
To the right: the nesting tree $X(f)$ corresponding to a fragment of the nodal picture
in Figure \ref{fig:sphere}, to the left, containing
$17$ nodal domains (where we neglected some small ones lying next to the boundary).
Figure \ref{fig:sphere} is essential
for deciding which components merge on the sphere outside of the fragment.}
\label{fig:nesting tree}
\end{figure}

As mentioned above, Nazarov and Sodin have determined the asymptotic law for the
cardinality $|\mathcal{C}(f)|$ of $\mathcal{C}(f)$ as $T\rightarrow\infty$.
There is a positive constant $\beta_{n,\alpha}$ depending on $n$ and $\alpha$
(and not on $M$) such that, with probability tending to $1$ as $T\rightarrow\infty$,
\begin{equation}
\label{eq:|C(f)| beta c vol T^n}
|\mathcal{C}(f)|\sim \beta_{n,\alpha}\frac{\omega_{n}}{(2\pi)^{n}}\vol(M) T^{n},
\end{equation}
here $\omega_{n}$ is the volume of the unit $n$-ball.
We call these constants $\beta_{n,\alpha}$ the Nazarov-Sodin constants. Except for
$n=1$ when the nodal set is a finite set of points and \eqref{eq:|C(f)| beta c vol T^n}
can be established by the Kac-Rice formula
($\beta_{1,\alpha} = \frac{1}{\sqrt{3}}\cdot \sqrt{1+\alpha+\alpha^{2}}$), these numbers are not known explicitly.

In order to study the distribution of the topologies of $\Cc(f)$ and
$\Omega(f)$ and the graph $X(f)$ we need certain discrete spaces as well as
their one-point compactifications. Let $\widetilde{H(n-1)}\cup \{ \infty\}$ denote the one-point
compactification of the discrete countable set of diffeomorphism classes
of compact connected manifolds of dimension $(n-1)$. Similarly, let
$\widetilde{B(n)}\cup\{\infty\}$ be the one-point compactification of
discrete countable set of diffeomorphism classes of $n$-dimensional compact connected 
manifolds with boundary, and $\widetilde{\Tc}\cup\{\infty\}$ be the
one-point compactification of the (discrete countable) set of connected rooted finite
graphs (i.e. graphs together with a marked node, referred to as the ``root").
Note that each $c\in\Cc(f)$ and $\omega\in \Omega(f)$ clearly determine
points in $\widetilde{H(n-1)}$ and $\widetilde{B(n)}$, which we denote by $t(c)$
and $t(\omega)$ respectively.

To each
(or at least almost each) edge $c$ in $\Cc(f)=E(X(f))$ we associate an end $e(c)$
in $\widetilde{\Tc}$ as follows: Removing $c$ from $X(f)$ leaves
either two components or one component. The latter will happen asymptotically very rarely and in this
case we ignore this edge $c$ (or we could make an arbitrary definition for $e(c)$).
Otherwise the two components are rooted connected graphs and we define the end
$e(c)$ to be smaller (in size) of these two rooted graphs (again, the event
that they are of the same size is very rare and can be ignored). With these
spaces and definitions we are ready to define the key distributions (they are
essentially probability measures) on $\widetilde{H(n-1)}$, $\widetilde{B(n)}$
and $\widetilde{\Tc}$ by:
\begin{eqnarray}
\label{eq:muC,Omega,X def}
\mu_{\Cc(f)} = \frac{1}{|\Cc(f) |} \sum\limits_{c\in \Cc(f)}\delta_{t(c)}, \\
\mu_{\Omega(f)} = \frac{1}{|\Omega(f) |} \sum\limits_{\omega\in \Omega(f)}\delta_{t(\omega)} \\
\mu_{X(f)} = \frac{1}{|\Cc(f) |} \sum\limits_{c\in \Cc(f)}\delta_{e(c)},
\end{eqnarray}
where $\delta_{\xi}$ is a point mass at $\xi$. These measures give the distribution
of topologies of nodal sets, nodal domains and ends of nestings for our given $f$.

Our first result asserts that as $T\rightarrow\infty$ and for a typical $f$ in
$\Ec_{M,\alpha}(T)$, the above measures converge $w$-star to universal measures
which depend only on $n$ and $\alpha$ (and not on $M$). Let $H(n-1)$ consist of all elements
of $\widetilde{H(n-1)}$ which can be embedded in $\R^{n}$, $B(n)$ of
those elements of $\widetilde{B(n)}$ that can be embedded in $\R^{n}$, and let 
$\Tc$ is the set of all finite rooted trees.

\begin{theorem}
\label{thm:muC,Omega,X exist}
There are probability measures $\mu_{\Cc,n,\alpha}$, $\mu_{\Omega,n,\alpha}$ and $\mu_{X,n,\alpha}$
supported on $H(n-1)\cup\{ \infty\}$,
$B(n)\cup \{\infty\}$ and $\Tc\cup \{ \infty\}$ respectively, such that
for any given $H\in \widetilde{H(n-1)}$, $B\in\widetilde{B(n)}$
and $G\in \widetilde{\Tc}$ and $\epsilon>0$,
\begin{equation*}
\begin{split}
\prob \big\{ f\in\Ec_{M,\alpha}(T):\: \max \big(&\left|\mu_{\Cc(f)}(H)-\mu_{\Cc, n ,\alpha}(H) \right|,
\left|\mu_{\Omega(f)}(B)-\mu_{\Omega, n ,\alpha}(B) \right|, \\
&\left|\mu_{X(f)}(G)-\mu_{X,n,\alpha}(G) \right|
\big) > \epsilon   \big\} \rightarrow 0,
\end{split}
\end{equation*}
as $T\rightarrow \infty$.
\end{theorem}

While the above ensures the existence of a law of distribution for the topologies, it gives
little information about these universal measures. A central issue is the support of these measures
and in particular:
\begin{enumerate}

\item
\label{it: charge inf?}
Do any of the $\mu_{\Cc,n,\alpha}$, $\mu_{\Omega,n,\alpha}$, $\mu_{X,n,\alpha}$
charge the point $\infty$, that is, does some of the topology of $V(f)$ escape in the limit?

\item
\label{it: charge fin?}
Are the supports of these measures equal to $H(n-1)$, $B(n)$ and $\Tc$ respectively,
i.e. do these measures charge each singleton in these sets, positively?

\end{enumerate}

\underline{Remarks:} (i). We expect that the answer to \eqref{it: charge inf?}
is NO and to \eqref{it: charge fin?} is YES (see below).
If the answer to \eqref{it: charge inf?} is no, then these measures capture the
full distribution of the topologies and Theorem \ref{thm:muC,Omega,X exist} can be stated in the stronger form
\begin{equation*}
\begin{split}
\prob \big\{ f\in\Ec_{M,\alpha}(T):\: \max \left( D(\mu_{\Cc(f)},\mu_{\Cc, n ,\alpha}),
D(\mu_{\Omega(f)},\mu_{\Omega, n ,\alpha}),D(\mu_{X(f)},\mu_{X,n,\alpha})\right)>\epsilon
   \big\} \rightarrow 0,
\end{split}
\end{equation*}
as $T\rightarrow\infty$, where the discrepancy is defined by
\begin{equation*}
D(\mu_{\Cc(f)},\mu_{\Cc, n ,\alpha}) = \sup\limits_{F\subseteq \widetilde{H(n-1)}}
\left|\mu_{\Cc(f)}(F)-\mu_{\Cc, n ,\alpha}(F) \right|,
\end{equation*}
the supremum being over all finite subsets $F$, and similarly for the other discrepancies.

(ii). The answer to \eqref{it: charge fin?} is only problematic in the monochromatic
case $\alpha=1$ (see Section \ref{sec:outline proofs}).

(iii). Once \eqref{it: charge inf?} and \eqref{it: charge fin?} are answered the qualitative
universal laws for topologies are understood. To get quantitative information the only
approach that we know is to do Monte-Carlo (numerical) experiments (see below).

Our main Theorems resolve \eqref{it: charge inf?} and \eqref{it: charge fin?} in a number of cases.
As far as the topologies of $V(f)$ we have:

\begin{theorem}
\label{thm:no escape, support}
We have $\mu_{\Cc,n,\alpha}(\{ \infty\})=0,$ and the support of $\mu_{\Cc,n,\alpha}$ is equal to $H(n-1)$.
In other words there is no ``escape of topology": $\mu_{\Cc,n,\alpha}(H(n-1))=1$, and $\mu_{\Cc,n,\alpha}$
charges every point positively.
\end{theorem}

The proof of Theorem \ref{thm:no escape, support} is outlined in the next section except for the
statement that every point of $H(n-1)$ is charged in the case $\alpha=1$. The latter is established in
the recent note ~\cite{CS}.

For $n=3$, $H(2)$ consists of all orientable compact connected surfaces $S$ of genus $g(S)\in G=\{0,1,2,\ldots\}$,
that is we identify $H(2)$ with $G$. In this case we give a different
treatment of Theorem \ref{thm:no escape, support} which yields
a little more information.

\begin{theorem}
\label{thm:mu support n=3}
The measure $\mu_{\Cc,3,\alpha}$ is supported in $G$ and charges each $g\in G$. Moreover, the mean of
$\mu_{\Cc,3,\alpha}$ as a measure on $G$ is finite.
\end{theorem}

In dimension $n=2$, $H(1)$ is a point, namely the circle, and the measure $\mu_{\Cc(f)}$ is trivial.
However $B(2)$ consists of all planar domains $\omega$ and these are parameterized by
their connectivity $m(\omega)\in\N$ (simply connected $m=1$, double connected $m=2$, ...), 
that is we can identify $B(2)$ and $\N$.

\begin{theorem}
\label{thm:mu support n=2}
\begin{enumerate}

\item
\label{it:mu Omega support n=2}
We have
\begin{equation*}
\mu_{\Omega,2,\alpha}\left[\{ \infty\} \right]=0
\end{equation*}
and the support of $\mu_{\Omega,2,\alpha}$ is all of $\N$, moreover the mean
of $\mu_{\Omega,2,\alpha}$ is at most $2$ (as a measure on $\N$).

\item
\label{it:support tree ends n=2}
The support of $\mu_{X,2,\alpha}$ contains all of $\Tc$ (but we don't know if
$\mu_{X,2,\alpha}(\{\infty\})=0$).

\end{enumerate}
\end{theorem}

\underline{Applications:} The extreme values of $\alpha$, namely $0$ and $1$ are the most
interesting. The case $\alpha=1$ is the monochromatic random wave
(and also corresponds to random spherical harmonics) and it has been suggested
by Berry \cite{Berry 1977} that it models the individual eigenstates of the quantization
of a classically chaotic Hamiltonian. The examination of the count of nodal domains  (for $n=2$)
in this context was initiated by ~\cite{BGS}, and ~\cite{BS}, and the latter suggest some interesting 
possible connections to exactly solvable critical percolation models.

The law $\mu_{\Omega, 2, 1}$ gives
the distribution of connectivities of the nodal domains for monochromatic waves.
Barnett and Jin's numerical experiments ~\cite{BJ} give the following values for its mass on atoms.

\vspace{5mm}

\begin{tabular}{|c|c|c|c|c|c|c|c|c|c|}
\hline
connectivity & 1 & 2 & 3 & 4 & 5 &6 & 7 \\
\hline
$\mu_{\Omega,2,1}$ &.91171 &.05143 &.01322 &.00628 & .00364 &.00230 &.00159 \\
\hline
\end{tabular}

\vspace{3mm}

\begin{tabular}{|c|c|c|c|c|c|c|c|c|c|}
\hline
connectivity &8 &9 & 10 & 11 & 12 & 13 & 14\\
\hline
$\mu_{\Omega,2,1}$ &.00117 &.00090 &.00070 &.00058 &.00047 &.00039 & .00034  \\
\hline
\end{tabular}

\vspace{3mm}

\begin{tabular}{|c|c|c|c|c|c|c|c|c|c|}
\hline
connectivity &15 &16 &17 &18 & 19 & 20 & 21\\
\hline
$\mu_{\Omega,2,1}$ &.00030 &.00026 &.00023 &.00021 &.00018 &.00017 &.00016  \\
\hline
\end{tabular}

\vspace{3mm}

\begin{tabular}{|c|c|c|c|c|c|c|c|c|c|}
\hline
connectivity & 22 & 23 &24 & 25 & 26\\
\hline
$\mu_{\Omega,2,1}$ &.00014 & .00013 &.00012 &.000098 &.000097 \\
\hline
\end{tabular}

\vspace{5mm}

The case $\alpha=0$ corresponds to the algebro-geometric setting of a random real
projective hypersurface. Let $W_{n+1,t}$ be the vector space of real homogeneous
polynomials of degree $t$ in $n+1$ variables. For $f\in W_{n+1,t}$,
$V(f)$ is a real projective hypersurface in $\mathbb{P}^{n}(\R)$.
We equip $W_{n+1,t}$ with the ``real Fubini-Study" Gaussian coming from the inner
product on $W_{n+1,t}$ given by
\begin{equation}
\label{eq:<f,g> FS}
\langle f,\, g \rangle = \int\limits_{\R^{n+1}} f(x)g(x) e^{-|x|^{2}/2} dx
\end{equation}
(the choice of the Euclidian length $|x|$ plays no role ~\cite{Sa}).
This ensemble
is essentially $\Ec_{M,0}(t)$ with $M=(\mathbb{P}^{n}(\R),\sigma)$ the projective
sphere with its round metric (see ~\cite{Sa}). Thus the laws $\mu_{\Cc,n,0}$ describe
the universal distribution of topologies of a random real projective hypersurface
in $\mathbb{P}^{n}$ (w.r.t. the real Fubini-Study Gaussian).

If $n=2$ the Nazarov-Sodin constant $\beta_{2,0}$ is such that the random oval
is about $4\%$ Harnack, that is it has about $4\%$ of the maximal number of
components that it can have (~\cite{Na}, \cite{Sa}).
The measure $\mu_{\Omega,2,0}$ gives the distribution of the connectivities of the nodal
domains of a random oval. Barnett and Jin's Montre-Carlo simulation for these yields:

\vspace{5mm}

\begin{tabular}{|c|c|c|c|c|c|c|c|c|c|}
\hline
connectivity & 1 & 2 & 3 & 4 & 5 &6 & 7 \\
\hline
$\mu_{\Omega,2,0}$ &.94473 &.02820 &.00889 &.00437 & .00261 &.00173 &.00128 \\
\hline
\end{tabular}

\vspace{3mm}

\begin{tabular}{|c|c|c|c|c|c|c|c|c|c|}
\hline
connectivity &8 &9 & 10 & 11 & 12 & 13 & 14\\
\hline
$\mu_{\Omega,2,0}$ &.00093 &.00072 &.00056 &.00048 &.00039 &.00034 & .00029 \\
\hline
\end{tabular}

\vspace{3mm}

\begin{tabular}{|c|c|c|c|c|c|c|c|c|c|}
\hline
connectivity &15 &16 &17 &18 & 19 & 20 & 21\\
\hline
$\mu_{\Omega,2,0}$ &.00026 &.00025 &.00021 &.00019 &.00016 &.00014 &.00013  \\
\hline
\end{tabular}

\vspace{3mm}

\begin{tabular}{|c|c|c|c|c|c|c|c|c|c|}
\hline
connectivity & 22 & 23 &24 & 25 & 26\\
\hline
$\mu_{\Omega,2,0}$ &.00011 & .00011 &.00009 &.00008 &.00008 \\
\hline
\end{tabular}

\vspace{5mm}

From these tables it appears that the decay rates of $\mu_{\Omega,2,1}(\{m \})$ and
$\mu_{\Omega,2,0}(\{m \})$ for $m$ large are power laws $m^{-\beta}$, with $\beta$
approximately $2.149$ for $\alpha=1$ and $2.057$ for $\alpha=0$. These are close to the universal
Fisher constant $187/91$ which governs related quantities in critical percolation ~\cite{KZ}.

The measure $\mu_{\Cc,3,0}$ gives the law of distribution of the topologies
of a random real algebraic surface in $\mathbb{P}^{3}(\R)$. It would be very interesting
to Monte-Carlo this distribution and get some quantitative information beyond Theorem
\ref{thm:mu support n=3}.

\begin{remark}
\label{rem:P-val top invar}
A $P$-valued topological invariant $F$ is a map $F:H(n-1)\rightarrow P$
(everything discrete). One defines the $F$-distribution of $f\in \Ec_{M,\alpha}(T)$ to be
\begin{equation*}
\mu_{F(f)} = \frac{1}{|\Cc(f)|}\sum\limits_{c\in\Cc(f)} \delta_{F(c)},
\end{equation*}
$\mu_{F(f)}$ is the pushforward of $\mu_{\Cc(f)}$ to $P$.
According to Theorem \ref{thm:no escape, support}, for the typical $f$, $\mu_{F(f)}$ will be close (in terms of
discrepancy) to the universal measure $\mu_{F,n,\alpha}$ on $P$, where
\begin{equation*}
\mu_{F,n,\alpha}(Y) = \mu_{Cc,n,\alpha}(F^{-1}(Y))
\end{equation*}
is the pushforward of $\mu_{\Cc,n,\alpha}$ to $P$.

For example, let $$F(c)=\text{Betti}(c) = (\beta^{(1)}(c),\,\ldots,\, \beta^{(k)}(c))$$
in $P_{n}:= (\Z_{\ge 0})^{k}$, where $n= 2k$ or $2k+1$ with $k>0$, and $\beta^{(j)}(c)$ is the $j$-th Betti
number of $c$ (the other Betti numbers are determined from the connectedness of $c$
and Poincare duality). From the fact that the support of $\mu_{\Cc,n,\alpha}$ is $H(n-1)$
one can show that $\mu_{\text{Betti},n,\alpha}$ is a (probability) measure on $P_{n}$ with full
support if $n$ is odd and with support $(\Z_{\ge 0})^{k-1}\times (2\Z_{\ge 0})$ if $n$ is even.
Moreover, the ``total Betti number"
\begin{equation}
\label{eq:tot Betti number}
\sum\limits_{y\in P_{n}} \left(\sum\limits_{j=1}^{k}y_{j}\right)\mu_{\text{Betti},n,\alpha}(\{y \})
\end{equation}
is finite.
In particular $\mu_{\text{Betti},n,0}$ describes the full law of distribution of the vector
of Betti numbers of a random real algebraic hypersurface in projective space.

\end{remark}

\subsection{Acknowledgements}
We would like to thank Mikhail Sodin
for sharing freely early versions of his work with Fedor Nazarov and
in particular for the technical discussions with one of us (Wigman) in Trondheim 2013,
and Ze\'{e}v Rudnick for many stimulating discussions.
In addition I.W. would like to thank Dmitri Panov and Yuri Safarov for sharing his expertise on
various topics connected with the proofs.
We also thank Alex Barnett for carrying out the numerical experiments
connected with this work and for his figures which we have included, as well
as P. Kleban and R. Ziff for formulating and examining the ``holes of
clusters" in percolation models. Finally, we thank Yaiza Canzani and Curtis McMullen
for their valuable comments on drafts of this note.

\section{Outline of proofs}

\label{sec:outline proofs}

\subsection{The covariance function for $\Ec_{\alpha}(T)$}
Most probabilistic calculations with the Gaussian ensemble $\Ec_{\alpha}(T)$ (we fix $M$)
start with the covariance function (also known as {\em covariance kernel})
\begin{equation}
\label{eq:K alpha covar def}
K_{\alpha}(T;x,y) := \E_{\Ec_{\alpha}}[f(x)f(y)] = \sum\limits_{\alpha T \le t_{j}\le T}\phi_{j}(x)\phi_{j}(y)
\end{equation}
(with suitable changes if $\alpha=1$). The function $K_{\alpha}$ is the reproducing kernel for our $\alpha$-band
limited functions. Note that
\begin{equation*}
\int\limits_{M}K_{\alpha}(T;x,x)d\nu(x) = \dim\Ec_{\alpha}(T) := \vol(M)D_{\alpha}(T),
\end{equation*}
where $D_{\alpha}(T)$ is the normalized dimension. The behaviour of $K_{\alpha}$ as $T\rightarrow\infty$ is decisive in
the analysis and it can be studied using the wave equation on $M\times \R$ and constructing a smooth parametrix
for the fundamental solution as is done in ~\cite{Lax,Horm}, see ~\cite{LPS} for a recent discussion.

Let
\begin{equation*}
\widetilde{K_{\alpha}}(T;x,y) = \frac{1}{D_{\alpha}(T)}K_{\alpha}(T;x,y),
\end{equation*}
then uniformly for $x,y\in M$,
\begin{equation}
\label{eq:K sim B(td(x,y))}
\widetilde{K_{\alpha}}(T;x,y) = \begin{cases}
B_{n,\alpha}(Td(x,y))+O(T^{-1}) &\text{if } d(x,y)T\ll 1\\
O(T^{-1}) &\text{otherwise}
\end{cases},
\end{equation}
where $d(x,y)$ is the distance from $x$ to $y$ in $M$, and for $w\in\R^{n}$
\begin{equation}
\label{eq:Balpha(w) def}
B_{n,\alpha}(w) = B_{n,\alpha}(\|w\|) = \frac{1}{|A_{\alpha}|} \int\limits_{A_{\alpha}}
e(\langle w,\xi\rangle) d\xi
\end{equation}
and $$A_{\alpha} = \{ w:\: \alpha \le \| w\| \le 1  \}.$$ Moreover, the derivatives of the
left hand side of \eqref{eq:K sim B(td(x,y))} are also approximated by the corresponding derivatives of the right hand
side.

Thus for points $y$ within a neighbourhood of $1/T$ of $x$ the covariance is given by
\eqref{eq:Balpha(w) def} while if $y$ is further away the correlation is small. This is
the source of the universality of the distribution of topologies, since the quantities we
study are shown to be local in this sense.

Let $H_{n,\alpha}$ be the infinite dimensional
isotropic (invariant under the action of the group of rigid motions, i.e. translations and rotations) Gaussian ensemble
(``field") defined on $\R^{n}$ as follows:
\begin{equation*}
f(x) = \sum\limits_{j=1}^{\infty}c_{j}\widehat{\psi_{j}}(x),
\end{equation*}
where $c_{j}$'s are i.i.d. standard (mean zero unit variance) Gaussian variables, and
$\psi_j$ are
an orthonormal basis of $L^2 (A_{\alpha}, dv )$,where $dv$ is the normalized
Haar measure. The covariance function
of $H_{n,\alpha}$ is given by
\begin{equation*}
\E_{H_{n,\alpha}}[f(x)f(y)] = B_{n,\alpha}(x-y).
\end{equation*}
The typical element in the ensemble $H_{n,\alpha}$ is $C^{\infty}$,
and the action by translations on $H_{n,\alpha}$ is ergodic by the classical Fomin-Grenander-Maruyama theorem
(see e.g. ~\cite{Gr}). As in ~\cite{So}
we show that the probability distributions that we are interested in are encoded in
this ensemble $H_{n,\alpha}$.

\subsection{On the proof of existence of limiting measures (Theorem \ref{thm:muC,Omega,X exist})}

For the existence of the measures in Theorem \ref{thm:muC,Omega,X exist} we follow the method
in ~\cite{NS} and \cite{So} closely. They examine the expectation and fluctuations of
the (integer valued) random variable $N(f,T)$ on $\Ec_{\alpha}(T)$ which counts the number
of connected components $c$ of $V(f)$. We examine the refinements of these given as:
for $$S\in H(n-1),$$ $N(f,S,T)$ is the (integer valued) random variable which counts
the number of such components $c$ which are topologically equivalent to $S$; for
$\omega\in B(n)$, $N(f,\omega,T)$ counts the number of components $c$ whose `inside' is homeomorphic
to $\omega$, and for $e\in \Tc$ a rooted tree $N(f,e,T)$ counts the number of
components $c$ whose end is $e$. The fact that our random variables are all dominated pointwise
by $N(f,T)$ allows us for the most part to simply quote the bounds for rare events developed
in ~\cite{So}, and this simplifies our task greatly. The basic existence result for each of our random
variables is the following, which we state for $N(f,S,T)$:
There is a constant $\widetilde{\mu_{n,\alpha}}(S)\ge 0$ such that
\begin{equation}
\label{eq:E[N(S)/T^n]->tilde{mu}}
\lim\limits_{T\rightarrow\infty} \E_{\Ec_{\alpha}(T)}\left[\left| \frac{N(f,S,T)}{T^{n}}-
\widetilde{\mu_{n,\alpha}}(S)  \right| \right] = 0.
\end{equation}

\vspace{3mm}

The constant $\widetilde{\mu_{n,\alpha}}(S)$ is determined from the Gaussian $H_{n,\alpha}$ as follows:
For $f\in H_{n,\alpha}$ and $R\ge 1$ let $N(f,S,R)$ be the number of components of $V(f)$ which
are homeomorphic to $S$ and which lie in $B(R)$ the ball about $0$ of radius $R$. This function
of $f$ is in $L^{1}$, and after suitable generalizations\footnote{The case of tree ends is the most subtle.}
of the sandwich estimates (~\cite{So}, page $6$)
for our refined variables one shows that the following limit exists and yields
$\widetilde{\mu_{n,\alpha}}(S)$:
\begin{equation*}
\widetilde{\mu_{n,\alpha}}(S)= \lim\limits_{R\rightarrow\infty} \frac{1}{\vol(B(R))}
\E_{n,\alpha}[N(f,S,R)].
\end{equation*}
Hence in terms of $\mu$,
\begin{equation*}
\mu_{\Cc,n,\alpha}(S) = \frac{\widetilde{\mu_{n,\alpha}}(S)}{\beta_{n,\alpha}\omega_{n}(2\pi)^{-n}}.
\end{equation*}

The proof of \eqref{eq:E[N(S)/T^n]->tilde{mu}} is in two steps. The first is a localization
in which one scales everything by a factor of $T$ in $1/T$-neighbourhoods of points in $M$,
and reduces the problem to that of the limit ensemble $H_{n,\alpha}$.
This process, called ``coupling" in ~\cite{So}, can be carried out in a similar way for $N(f,S,T)$
(as well as our other counting variables) after relativizing the various arguments
and inequalities. The second step concerns the study of the random variable
$N(f,S,R)$ (and again the other counts) on $H_{n,\alpha}$, asymptotically
as $R\rightarrow\infty$. A key point is that
this latter variable is firstly measurable (it is locally constant) and is in
$L^{1}(H_{n,\alpha})$. As in ~\cite{So} this allows one to apply the ergodic
theorem for the group of translations of $\R^{n}$ to ensure that the counts
in question converge when centred at different points. This provides the `soft'
existence for the limits at hand while providing little further information.
As a ``by-product" this approach implies that a typical nodal
domain or a tree end of $f$ lies in a geodesic ball of radius $R/T$ in $M$ for $R$
large (but fixed); this ``semi-locality" is the underlying reason for the ergodic
theory to be instrumental for counting nonlocal quantities.

\vspace{3mm}

As was mentioned above, to infer information on $f$ from $H_{n,\alpha}$ one needs
to construct a coupling, that is, a copy of $H_{n,\alpha}$ defined on the same probability space
as $f$, so that with high probability a random element $g$ of $H_{n,\alpha}$
is merely a small perturbation of (the scaled version of) $f$ in $C^{1}(\overline{B(2R)})$
(that is, both the values and the partial derivatives of $g$ approximate
those of $f$); this is possible thanks to \eqref{eq:K sim B(td(x,y))}.
Moreover, in this situation, with high probability both $f$ and $g$ are ``stable",
i.e. the set of points where both $f$ and $\| \nabla f\|$ are small
is {\em negligible} (the same holding for $g$).
Nazarov and Sodin used the ingenious ``nodal trap" idea, showing that each of the nodal components
$c$ of $g$ is bounded between the two hypersurfaces $g^{-1}(\pm \epsilon)$, to prove that
under the stability assumption $c$ corresponds to a {\em unique}
nodal component of $f$. This allowed them to infer that the nodal count of $g$
is approximating the nodal count of $f$ (neglecting the unstable regions). We refine their
argument by observing that the topological class of a nodal component inside the
``trap" cannot change while perturbing from $g$ to $f$, as otherwise, by Morse Theory, one would have
encountered a low valued critical point; this is readily ruled out by
the stability assumption. The same approach shows that neither the diffeomorphism class
of the corresponding nodal domain nor the local configuration graph can change
by such a perturbation. This completes the outline of the proof of Theorem \ref{thm:muC,Omega,X exist}.

\subsection{The measures $\mu_{\Cc,n,\alpha}$ do not charge $\infty$}

To establish the claims in Theorems \ref{thm:no escape, support} and \ref{thm:mu support n=3}
concerning the supports of the measures $\mu_{\Cc,n,\alpha}$
(respectively $\mu_{\Omega,n,\alpha}$, $\mu_{X,n,\alpha}$)
one needs to input further topological and analytic arguments.
The first part of Theorem \ref{thm:no escape, support} is deduced from uniform upper bounds for the
mass of the tails of the measures on $H(n-1)$ which approximate $\mu_{\Cc,n,\alpha}$.
For the Gaussian field $H_{n,\alpha}$ this involves controlling the topologies of most
(i.e. all but an arbitrarily small fraction) of the components $c$ of $V(f)$ in balls of
large radius $R$. Here $f$ is typical in $H_{n,\alpha}$, and the control needs to be
uniform in $R$.

From the Kac-Rice formula, which gives the number of critical points of $f$, one can show that
most components $c$ arise from a bounded number of surgeries starting from $\mathcal{S}^{n-1}$.
Hence, by Morse theory, this is enough to bound the Betti numbers of $c$; however, a priori
the topology of $c$ could lie in infinitely many types. To limit these we examine
the geometry of $c$ (in the induced metric from $\R^{n}$). The key is a uniform bound
for the derivatives of the unit normal vector at each point of $c$ (for most
components). This is achieved by extending arguments in ~\cite{NS} to typical $f$'s using
among other things the Sobolev embedding theorem. Once we have uniform bounds for the volume,
diameter and sectional curvatures on $c$, Cheeger's finiteness theorem (~\cite{Ch,Pe}) ensures that $c$
lies in only finitely many diffeomorphism types.

The low dimensional cases not charging $\{\infty\}$ in Theorems \ref{thm:mu support n=3} and \ref{thm:mu support n=2}
are approached more directly.
For part \eqref{it:mu Omega support n=2} of Theorem \ref{thm:mu support n=2}
take $M=\mathcal{S}^{2}$ with its round metric. For almost all $f\in \Ec_{M,\alpha}(T)$, $V(f)$
is nonsingular and the graph $X(f)$ is a tree (by the Jordan curve Theorem).
Hence, if $d(\omega)$ is the degree of the vertex $\omega$, then
$$\sum\limits_{\omega\in \Omega(f)}d(\omega) = 2|\Omega(f)|-2,$$
and the mean (over $\N$) of $\mu_{\Omega}(f)$ is equal to $2-\frac{2}{|\Omega(f)|}$.
It follows that the limit measures $\mu_{\Omega,2,\alpha}$ do not charge $\{\infty\}$,
and that their means are at most $2$. From the data in Barnett's tables (see Section $1$)
it appears that the means for $\alpha=0$ and $\alpha=1$ may well be less than $2$. If
this is so it reflects a nonlocal feature of ``escape of topology" at this more quantitative
level.

The proof of Theorem \ref{thm:mu support n=3} 
uses the Kac-Rice formula (see ~\cite{Cramer-Leadbetter,Adler-Taylor}) and some topology. The expected value
of the integral of over $M$ of ``any" local quantity, such as the curvature $\kappa(x)$
of the surface $V(f)$ ($n=3$), over $\Ec_{M,\alpha}(T)$ can be computed. For example,
if $M=\R\mathbb{P}^{3}$ with its round Fubini-Study metric, then by Gauss-Bonnet
\begin{equation*}
\E_{\Ec_{M,\alpha}(T)} \left[ \sum\limits_{c\in \Cc(f)}2(1-g(c)) \right] \sim -\gamma_{\alpha}T^{3},
\end{equation*}
where $g(c)$ is the genus of $c$
(here $\gamma_{\alpha}>0$ and for $\alpha=0$ it is computed explicitly in ~\cite{Bu}). Hence
as $T\rightarrow\infty$,
\begin{equation*}
\E_{\Ec_{M,\alpha}(T)} \left[ \sum\limits_{c\in \Cc(f)}g(c)\right] \sim \left(\gamma_{\alpha}+
\frac{2\vol(\R\mathbb{P}^{3})\beta_{3,\alpha}\omega_{3}}{(2\pi)^{3}}\right)\cdot T^{3}.
\end{equation*}
From this one deduces that $\mu_{\Cc,3,\alpha}(\{\infty\}) =0$, and that the mean of
$\mu_{\Cc,3,\alpha}$ (over $G$) is at most $$2+\frac{\gamma_{\alpha}}{\vol(\R\mathbb{P}^{3})
\cdot \beta_{3,\alpha}\omega_{3}(2\pi)^{-3}}.$$

\subsection{The measures $\mu_{\Omega,2,\alpha}$ and $\mu_{\Cc,3,\alpha}$
charge every finite atom}

The proof that the measures $\mu_{\Cc,n,\alpha}$, $\mu_{\Omega,n,\alpha}$ and $\mu_{X,n,\alpha}$
charge every topological atom reduces to producing an $f\in H_{n,\alpha}$ for which the
corresponding $V(f)$ (respectively $\Omega(f)$, $X(f)$) has the sought atomic configuration
(since the $g$'s in a suitably small $C^{k}$ neighbourhood of $f$ have the same local
configuration level and such a neighbourhood has positive measure in $H_{n,\alpha}$).
For $\Sigma\subseteq \R^{n}$ a compact set let
\begin{equation*}
\widehat{\Sigma} = \left\{   f:\: f(x) = \sum\limits_{\xi\in\Sigma} a_{\xi}e(\langle x,\xi\rangle),
\text{with } a_{\xi}=0 \text{ for all but finitely many } \xi \right\}.
\end{equation*}

For our purposes it suffices to find a $g\in\widehat{A_{\alpha}}$ (real valued) with the desired
topological atom. If $0\le \alpha < 1$ then one can show that for any compact ball $B$,
$\widehat{A_{\alpha}}|_{B}$ is dense in $C^{k}(B)$ (for any $k$). Hence constructing
a function $g$ of the type that we want is straightforward. However for $\alpha=1$, the
closure of $\widehat{A_{1}}|_{B}$ in $C(B)$ is of infinite codimension.
Nevertheless,
the following much weaker statement holds:

\begin{figure}[ht]
\centering
\includegraphics[height=30mm]{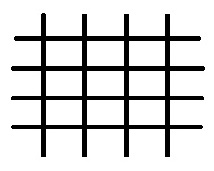}
\caption{A grid, the nodal set of $f_{0}$.}
\label{fig:grid}
\end{figure}

\begin{lemma}
\label{lem:K finite val no rest}
For $n\ge 2$ and $K\subseteq \R^{n}$ a finite set, $\widehat{A_{1}}|_{K}=C(K)$, i.e. there
is no restriction on the values attained by a function in $\widehat{A_{1}}$ on a finite set.
\end{lemma}

\begin{figure}[ht]
\centering
\includegraphics[height=60mm]{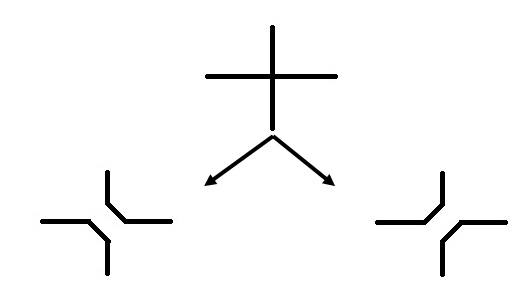}
\caption{A singularity resolution.}
\label{fig:singularity resolution}
\end{figure}

Our proof uses asymptotics of Bessel functions and $L^{\infty}$ bounds for spherical harmonics.
One can also deduce Lemma \ref{lem:K finite val no rest} for $K$'s which are subsets of $\Z^{n}$
(and this is sufficient for our purposes)
from Ax's ``function field Schanuel Theorem" \cite{Ax}. In fact, one
can deduce a much more general result which is useful in this context:
If $\Sigma$ is an $r$-dimensional ($r\ge 1$) real algebraic subvariety
of $\R^{n}$ which is not special in the sense of ~\cite{Pila}, and
$K\subseteq \Z^{n}$ is finite, then $\widehat{A_{1}}|_{K}=C(K)$.

\begin{figure}[ht]
\centering
\includegraphics[height=55mm]{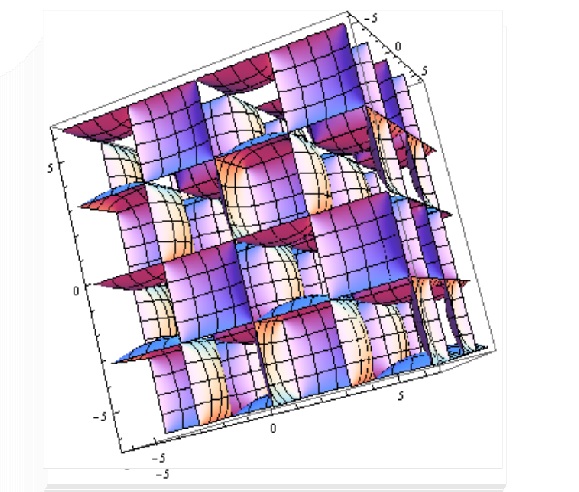}
\includegraphics[height=55mm]{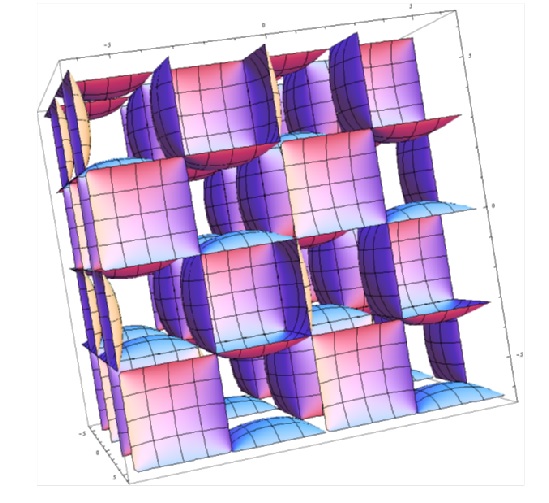}
\caption{A fragment of the zero set of $$f_{0}(x_{1},x_{2},x_{3})=\sin{x_{1}}\sin{x_{2}}+\sin{x_{1}}\sin{x_{3}}+
\sin{x_{2}}\sin{x_{3}}$$ shown from two different perspectives.
It consists of a lattice arrangement of infinitely many layers each containing boxes arranged on half
of the black cells of an infinite chessboard, diagonally connected to the upper and lower layers.}
\label{fig:smooth boxes}
\end{figure}

To produce an $f\in \widehat{A_{1}}$ (for $n=2$) with $X(f)$ having a given end $e\in \Tc$,
start with $$f_{0}(x_{1},x_{2})=\sin(\pi x_{1})\sin(\pi x_{2}),$$ whose nodal set $V(f)$
is a grid (see Figure \ref{fig:grid}) with conic singularities at the points of $\Z^{2}$.
For any finite $K\subseteq \Z^{2}$ we can choose $\psi(x_{1},x_{2})$
in $\widehat{A_{1}}$ with $\psi(k)=\epsilon_{k}\in \{-1,1\}$ for $k\in K$,
where $\epsilon_{k}$ is any assignment of signs. Set
\begin{equation*}
f(x_{1},x_{2}) = f_{0}(x_{1},x_{2})+\epsilon\psi(x_{1},x_{2}),
\end{equation*}
where $\epsilon$ is a small positive number. The singularity at $k$ will resolve in either
of the forms as in Figure \ref{fig:singularity resolution}, according to the sign of $\epsilon_{k}$.
One shows that this gives enough flexibility by choosing $K$ and $\epsilon_{k}$ to produce
any rooted tree in $\Tc$. This completes the outline of Theorem \ref{thm:mu support n=2},
part \eqref{it:support tree ends n=2}.

\vspace{5mm}

For a proof of Theorem \ref{thm:mu support n=3} 
one uses the above Lemma in a similar way starting with the function
\begin{equation*}
f_{0}(x_{1},x_{2},x_{3})= \sin(\pi x_{1})\sin(\pi x_{2})+\sin(\pi x_{1})\sin(\pi x_{3})
+\sin(\pi x_{2})\sin(\pi x_{3}).
\end{equation*}
The singularities of $V(f_{0})$ are at integral lattice points and are conic
(see Figure \ref{fig:smooth boxes}).
Perturbing $f_{0}$ near such a point $k$ resolves $V(f_{0})$ to a $1$-sheeted or
$2$-sheeted hyperboloid depending on the sign of $\epsilon_{k}$. Again, one shows by examining
the components of $\R^{3}\setminus V(f_{0})$ (which consists of infinitely many alternating
cubes, and the complement, which is connected), that perturbing $f_{0}$
by a suitable $\epsilon \psi\in \widehat{A_{1}}$ is enough to produce any element of
$H(2)$ as a component of $V(f)$. 

\begin{figure}[ht]
\centering
\includegraphics[height=60mm]{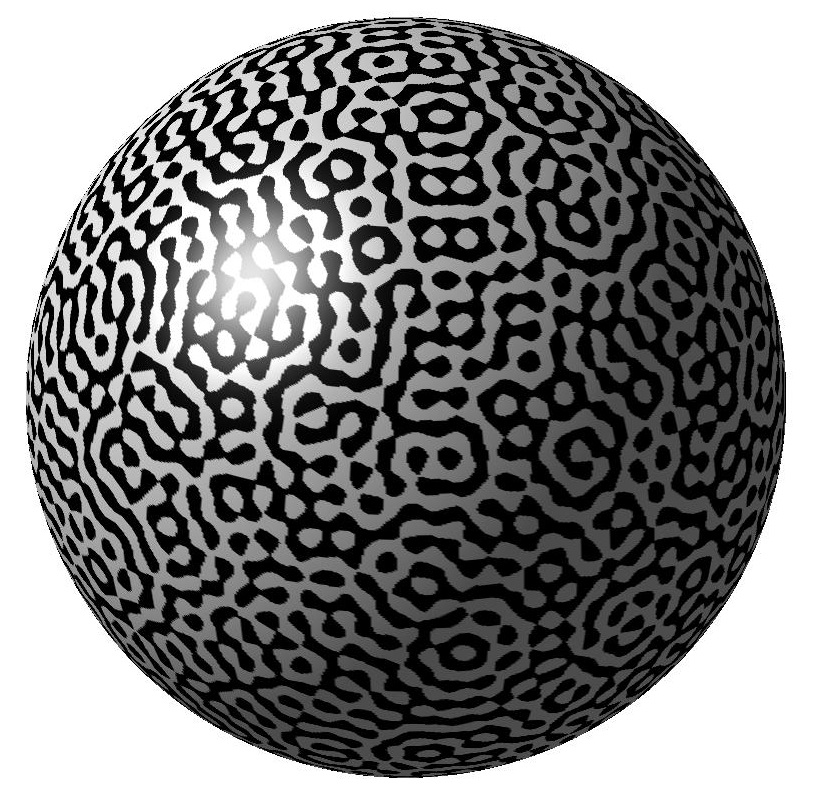}
\includegraphics[height=60mm]{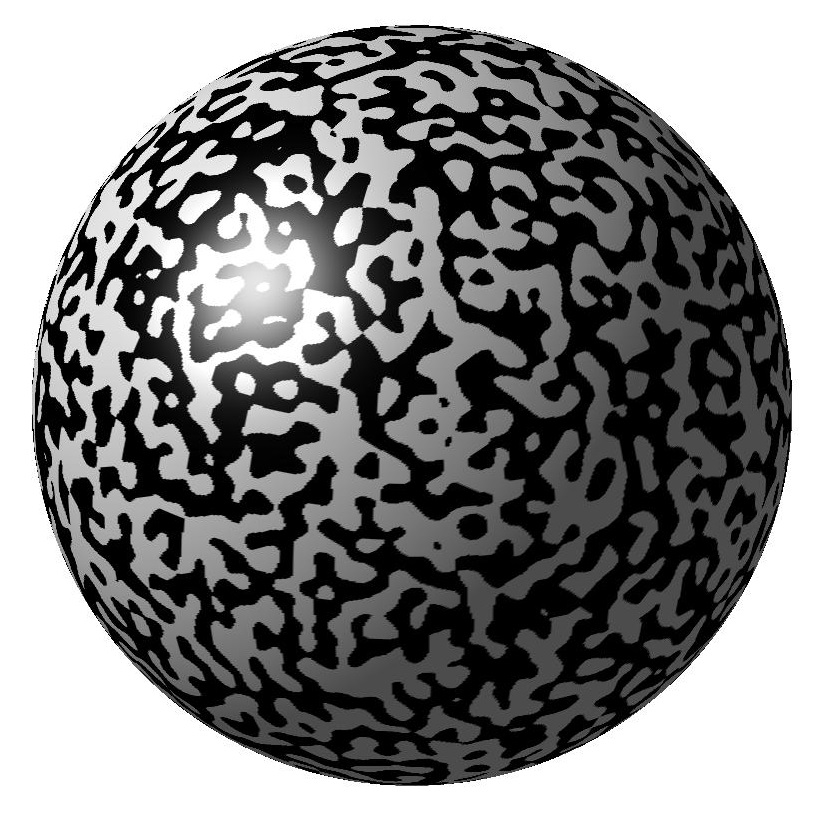}
\caption{Nodal line and domains for a random element in
$\Ec_{\mathcal{Sc}^{2},\alpha}$ with $\alpha =1$ (left) and $\alpha =0$ (right),
$T=\sqrt{80\cdot 81}$, pictures produced by Alex Barnett.
The nodal domains are the black and white connected components,
and the nodal line is the interface
between these.}
\end{figure}

\vspace{10mm}

We end with comments on Remark \ref{rem:P-val top invar}. According to ~\cite{Mi}, the total
Betti number $\sum\limits_{j=1}^{n-1}\beta^{j}$ of the zero set of a nonsingular real homogeneous
polynomial of degree $t$ in $(x_{0},x_{1},\ldots,x_{n})$ is at most $t^{n}$. This together with
\eqref{eq:|C(f)| beta c vol T^n} implies the finiteness assertion \eqref{eq:tot Betti number}
which in turn ensures that
$\mu_{\text{Betti},n,0}(\{\infty\}) = 0$. That the image of $H(n-1)$ under Betti is restricted
as claimed follows from our $c$'s bounding a compact $n$-manifold so that $\chi(c)$ is even.
On the other hand, starting from $\mathcal{S}^{n-1}$ and applying suitable $p$-surgeries which increase
$\beta^{p+1}$ by $1$ if $p+1$ is not in the middle dimension and by $2$ if it is, shows
that the image of Betti is as claimed. An interesting question about the Betti numbers
raised in ~\cite{GW} page $4$ in the context of their ensembles, is whether the limits
\begin{equation*}
\frac{1}{\beta_{n,0}T^{n}}\E_{\Ec_{\mathbb{P}^{n},0}(T)}[\beta^{j}(V_{f})], \; T\rightarrow\infty
\end{equation*}
exist for each $1\le j\le k$? If so a natural question is whether these are equal to the corresponding
mean for $\mu_{\text{Betti},n,0}$? These appear to be subtle questions related to the possible
non-locality of these quantities (escape of mass) and it is unclear to us what to expect.

\end{document}